\documentclass[conference,a4paper]{IEEEtran}
\IEEEoverridecommandlockouts
\usepackage{cite}
\usepackage{amsmath,amssymb,amsfonts}
\usepackage{algorithmic}
\usepackage{graphicx}
\usepackage{textcomp}
\usepackage{xcolor}
\def\BibTeX{{\rm B\kern-.05em{\sc i\kern-.025em b}\kern-.08em
    T\kern-.1667em\lower.7ex\hbox{E}\kern-.125emX}}

\usepackage{subcaption}

\begin{document}

\title{Modelling a CubeSat-based Space Mission and its Operation\\
}
\author{
\IEEEauthorblockN {Carlos L. G. Batista\IEEEauthorrefmark{1},
F\'atima Mattiello-Francisco\IEEEauthorrefmark{1}
\IEEEauthorblockA{\IEEEauthorrefmark{1}
National Institute for Space Research (INPE)\\
Space Systems Engineering\\
S\~ao Jos\'e dos Campos, Brazil\\
Email: \{carlos.batista, fatima.mattiello\}@inpe.br}
}}

\maketitle

\begin{abstract}
Since the early 2000’ years, the CubeSats have been growing and getting more and more “space'' in the Space industry. Their short development schedule, low cost equipment and piggyback launches create a new way to access the space, provide new services and enable the development of new technologies for processes and applications. That is the case of the Verification and Validation of these missions. As they are cheaper to launch than traditional space missions, CubeSats win by numbers. With more than 1000 CubeSats launched they still achieve less than 50\% rate of successful missions and that is caused mainly by poor V\&V processes. Model Based approaches are trying to help in these problems as they help software developers along the last years. As complex systems, space products can be helped by the introduction of models in different levels. Operational goals can be achieved by modeling behavioral scenarios and simulating operational procedures. Here, we present a possible modeling solution using a tool that integrates the functionalities of FSM and Statechartes, the ATOM SysVAP (System for Validation of Finite Automatons and Execution Plans). With this tool we are able to model the behaviour of a space mission, from its top level (i.e. system and segments) to its low level (subsystems) and simulate their interactions (operation). With the help of Lua Programming Language, it is possible to generate analysis files, specific scenarios and control internal variables.
\end{abstract}


\section{Introduction}

The innovation through the micro/nano/picosatellites has been creating a new whole market and habitat for the space sector. With a short life cycle, low cost and the use of standardized platforms, these satellites enable the in-orbit qualification of new space technologies, high science and 
human resources \cite{martin2015newspace}. The CubeSat standard , a.k.a. U-class, has predefined mechanical and electrical interfaces, including launcher interfaces \cite{heidt2000cubesat}. They brought a new satellite development philosophy, not only for the university satellites but also for all the small-size ones. More than 1300 nanosatellites have been launched in the last fifteen years, 1200 of them are CubeSats and the foreseeable future expect more than 6 thousand in the next six years \cite{villela2019towards}.

The real deal is that in order to develop these missions in universities, reliability requirements, usually specified at traditional space missions, were reduced. Therefore, these satellites have demonstrated high failure rate.
Some recent studies \cite{villela2019towards,swartwout2016secondary} pointed out that 50\% of the CubeSat-based missions fail, and 70\% of them are developed by inexperienced teams.
The big difference between the traditional developers and these “hobbyists” is the lack of good practices in design, assembly, and tests. We still have a lack of studies showing us the main problems that cause these missions to fail, especially during the operational phase. For most of them the absence of processes and procedures documentation is the main problem to investigate this kind of operation failure \cite{batista2020impacts}.

In this context we can point out the lack of testing methods and operational procedures validated in ground, which are techniques used in the satellite development cycle to support the verification of the expected behavior of the system in the operational environment, increasing the system reliability \cite{berthoud2019university}.

The Verification and Validation -- V\&V -- processes are crucial to all space missions to be successful. But for most traditional space missions they are usually onerous in terms of time and money costs, which is incompatible with the U-class development philosophy, "faster and cheaper" \cite{forsberg19994}.

The fundamental for any V\&V processes is the system requirements specification. Most CubeSat-based projects have given less importance to the mission requirements definition and analysis phase because the use of the CubeSat standard is considered the architectural solution to the satellite platform. Although CubeSat provides COTS (Commercial Off The-Shelf) subsystems with standardized interfaces and communication protocols,  they are not enough to define the mission goals. The mission requirements and constraints are the major elements to drive the use of the COTS in right way, as result of Phase 0 analysis, i.e. mission conception analysis and feasibility.

The mission concept of operation -- ConOps -- is a discipline in space system engineering that plays a very important role in the mission requirements analysis.
ConOps addresses the mission analysis under the user’s perspective. It traverses both ground and space segments, aiming to highlight dependencies among mission elements and the space environment to meet the mission operation requirements \cite{wertz1999smad}.
The mission ConOps analysis deals with multiple operation scenarios that reveal requirements and needed design functions. 

V\&V and ConOps are intimately related because the system requirements and specifications derived from the operational needs shall be verified and the intended use of the system shall be validated considering the stakeholders expectations \cite{nasa2016handbook,ecss:hb02a:verification}.


Modelling the expected behavior of a system for requirements analysis purpose has been helping different engineering areas across the years. Model Based approaches are getting more and more of this piece of the market, including Cube/NanoSats due to its speed, reliability, and reproducibility. Modelling allows to represent the system in different level of abstraction, under different perspective. Depending on the purpose, appropriated modelling formalism is used to explore particular aspect of the system. 

Focusing on the concept of operation aspects of a Cubesat-based mission, we present an experience on the use of a modelling tool developed at INPE, by space system engineering student.  The tool aids the mission design analysis allowing anticipate requirements verification by means of operational scenarios validation. The purpose is to increase confidence in the mission requirements, reducing the cost of design rework in terms of time and resources consuming.

\section{Related Works}

New initiatives have been trying to increase the chance of success of CubeSat-based missions using different approaches, mainly Model-Based and Model-Driven approaches. Modelling tools have helped many projects and missions to achieve the confidence on their projects design and development before going to production and operations since the beginning of space racing.

\cite{kaslow2018vevcrm} presents the CubeSat Reference Model -- CRM -- a reference example that is being developed by the INCOSE Space Systems Working Group -- SSWG. The goal of the CRM is to facilitate the design, verification and validation of a CubeSat project using SysML. The CRM is being developed flexible enough to be used in different missions and different teams.

Other studies, \cite{jacklin2015vvsurvey,khan2012model}, focus on Model Driven Engineering, aiming to automate the V\&V of on-board embedded software. The description of a MDV\&V, or Model Driven Verification and Validation, uses SysML to anticipate the platform V\&V, suggesting a systemic approach to simulate all the real spacecraft tests. And the usage of a new models philosophy (activity of space product V\&V where different models are specified for the space mission, like MockUps, Engineering Model, Qualification Model, Flight Model, etc) with intermediate models, e.g. FlatSats and DevSats \cite{eickhoff2003mdvv}.

Two highlighted works on this field are: (i) the development of a process and metamodel, in Arcadia Method, using Capella, for the flux of information and functionalities of an spacecraft from its ConOps in order to automatically fulfil the parametric information of a configuration file in a orbiter/operational simulator \cite{pallamin2019metamodel} and; (ii) the proposed Scalable Architecture Test System -- SATS --, in order to support the V\&V, combine MDE and MBT to anticipate the possible problems of interoperability and robustness \cite{paiva2019sats}.


\section{Modelling a Space Mission}

Aiming at representing the behavior of Cubesat-based mission under the perspective of operation for studies purpose concerning operation scenarios telecommanded from ground stations, the ATOM SysVAP modelling tool was used.  Four main reasons support the choice:  (i) the tool was developed in-house. ATOM SysVAP is a result from an INPE's master student work;  open source and available online in a git repository; (ii) this tool has been used in graduation class, at INPE, to aid students on the specification of Operational Procedures, visualizing their execution on an operational simulator (iii) the use of Lua Programming scripts creates a new level of modeling where it is possible to associate full operational scripts to different states and events and; (iv) the use of FSM is more simple and easy to deal and interfacing it with other modelling and model checking tools, if necessary.

\subsection{ATOM SysVAP}

The ATOM SysVAP -- Atom SYStem for Validation of finite Automatons and execution Plans --  is a tool developed to validate an architecture proposed to execute execution operational plans \cite{ivoandre2013atom}.

The ATOM enables the modeler to: (i) represent the Transition and States Domain as Mealy, Moore Machines and Statecharts; (ii) represent sub-FSM as part of other FSM; (iii) execute and validate operational plans in debug/real time mode and; (iv) access to environmental variables in order to evaluate the results and determine if the operational plan can or cannot be accepted.

Also, the ATOM uses the Lua Programming language for the development of the Domain Language. This capability supports the model execution offering new facilities for the data description.

\subsection{Elements of the Model}

\begin{figure}[!h]
    \centering
    \includegraphics[width=0.4\textwidth]{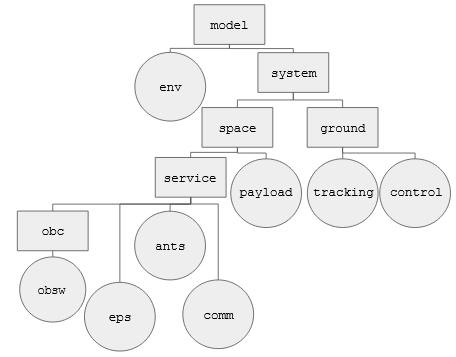}
    \caption{Model breakdown hierarchy}
    \label{fig:hierarchy}
\end{figure}

The element of the modelled space systems follows the hierarchical sequence presented at the Figure~\ref{fig:hierarchy}.

At Figure~\ref{fig:hierarchy}, the rounded shapes represent FSM without sub-FSM (bottom level of abstraction) as the square shapes represent FSM that can be grained in order to lower the level of abstraction.

From the top level, the FSM “model” represents the model as a whole, their sub-FSM are representations of what is inside the system to be modeled, “system”, and what is outside but still interferes with the system, the “environment”. As a space system, we are able to divide into two other sub-FSM, “space segment” and “ground segment”. This separation is common at space missions to more easily organize the attributions, needs and capabilities of each one.
At this point the ground segment can be divided into “tracking”, a model to deal with the pointing and passover constraints of ground stations and satellite, and “control”, the model responsible to deal with the operation of the satellite, this model can represent specific operational plans as multiple FSM. Similarly, the space segment can be divided into the “service” and “payload” buses. As a service bus, the platform is divided again into the different subsystems. As an example of granularity, the onboard computer model can be splitted into a sub-FSM to represent only  the logical functioning of the embedded software, “obsw”.

Considering the launcher requirements for CubeSats, (a) the satellite must be in “off” state while inside the dispenser; (b) the satellite must have a device to ensure the batteries are disconnected to the power system, a killswitch, while inside the dispenser and; (c) the satellite must not enter any operational state or boot mode for at least 30 minutes after the launch from the dispenser.

\begin{figure}
    \centering
    \begin{subfigure}[b]{0.1\textwidth}
        \centering
        \includegraphics[width=\textwidth]{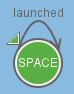}
        \caption{Space FSM}
        \label{fig:space}
    \end{subfigure}
    \begin{subfigure}[b]{0.35\textwidth}
        \centering
        \includegraphics[width=\textwidth]{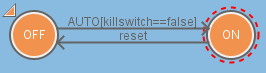}
        \caption{Electric Power Subsystem FSM}
        \label{fig:eps}
    \end{subfigure}
    \begin{subfigure}[b]{0.45\textwidth}
        \centering
        \includegraphics[width=\textwidth]{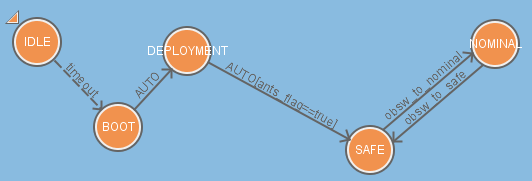}
        \caption{OnBoard Software FSM}
        \label{fig:obsw}
    \end{subfigure}
    \begin{subfigure}[b]{0.45\textwidth}
        \centering
        \includegraphics[width=\textwidth]{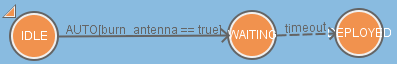}
        \caption{Antennas FSM}
        \label{fig:ants}
    \end{subfigure}
    \caption{Examples of modelled FSM}
    \label{fig:fsm_examples}
\end{figure}

So, the Figure~\ref{fig:fsm_examples} represents the models used for this situation, once the space segment receive the event “launched” (fig. 2a), the variable “killswitch” turns false, activating the event at the EPS -- Electric Power Subsystem -- model (fig 2b). The powering on of the EPS activates two events in two different FSM:  (i) the OBSw -- On Board Software -- FSM starts a countdown of 30 minutes to start the booting process, BOOT state, (fig 2c) and; (ii) at the Antennas Subsystem, the deployment process of the antennas is started also (fig.2d). Only after the 30 minutes, the OBSw is able to go to the DEPLOYMENT state, where it awaits for the state of the antennas (deployed or not, i.e. ants\_flag variable) changing its state to SAFE. DEPLOYMENT, SAFE and NOMINAL are states considered as operational for satellite already.

\subsection{Analysis Files}

Besides the modelling capabilities of the ATOM SysVAP tool, one great advantage and feature of the IDE is the integration with the Lua Programming Language. With that we are capable of reading and writing “comma separated values” files, a.k.a. csv files.

Once we have modelled the behaviour of the intended space system, some routines and functions in Lua can be used to generate analysis files, as csv files. These analysis files represent the internal variables that we want to monitorate, such as telemetries and more.

\begin{figure}[!ht]
    \centering
    \includegraphics[width=0.50\textwidth]{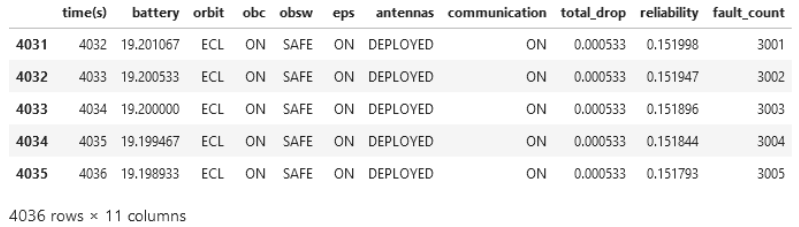}
    \caption{Example of analysis file opened in Jupyter Notebook}
    \label{fig:analysis}
\end{figure}

For example, Figure~\ref{fig:analysis} shows an analysis file, read by a python notebook, with information about the spacecraft battery voltage level, the subsystems FSM actual states, and even data about a reliability model implemented and the number of faults in the system, also implemented due to the Lua capabilities.

The generation of these analysis files can help the development team to check the behaviour of specific variables of the spacecraft and how to control them and also help the operational team to understand them even during the operational phase.

\begin{figure}
    \centering
    \begin{subfigure}[b]{0.4\textwidth}
        \centering
        \includegraphics[width=\textwidth]{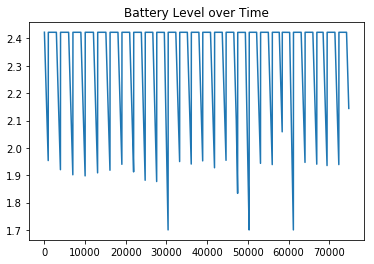}
        \caption{Battery Level}
        \label{fig:batt}
    \end{subfigure}
    \begin{subfigure}[b]{0.4\textwidth}
        \centering
        \includegraphics[width=\textwidth]{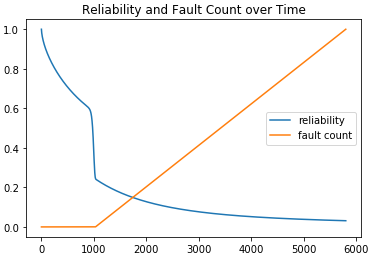}
        \caption{Reliability and Fault Count}
        \label{fig:rely}
    \end{subfigure}
    \caption{Examples of graphs derived from the analysis file}
    \label{fig:graphs}
\end{figure}

Figure~\ref{fig:graphs} helps us to understand some graphs that can be plotted in order to evaluate the behaviour of some variables. At Figure~\ref{fig:batt} we can access information about the battery level as each state of the environment changes (i.e. sun or eclipse) and if a certain number of subsystems are powered or not. As this, we can monitor the depth of discharge of the battery and simulate different conditions of operation taking into account the power budget. Figure~\ref{fig:rely} represents a derived comparison, from a fault model minimally implemented from a reliability model (a Weibull distribution). In this case, the chance of a fault occurrence, at the system level, is calculated as a normal distribution from 0 to 1. Once the chance of a fault is less than the reliability level at the same point in time, a fault occurs.

That is a simple fault model implemented without any fault behaviour associated with.

\section{Conclusions and Future work}

As we evolve the technologies for the access of space, the processes, procedures, activities and methods should evolve together.

We presented an experience on the use of a modelling tool, ATOM SysVAP, to represent the behavior of Cubesat-based mission, under operational perspective, for the purpose of  the verification and validation as a point of view from ConOps. The purpose is to highlight the dependencies among the major elements of the mission and space environment under perspective of the mission operation, which contributes to increase confidence on the definition of system requirements, reducing the project development cycle and the time spend during V\&V processes. 

The ATOM SysVAP has demonstrated, so far, its capability to, not only, model the operational phase of a space system, but also help the conception, development, and V\&V (validation of specific FSM paths). With the use of Lua as scripting language, the ATOM exploits its usage being capable of generating telemetry files, payload data, mathematical models, evaluating behaviour in real time (debug) and, modeling faults (using fault/error/failure behaviours and scenarios). This capability enables the FSM to be more reduced and the models simpler.


Moreover, the method used at this point is not completely tool independent. Based on the capabilities of the ATOM SysVAP, the models were evolved and implemented in order to represent the space mission elements and behaviour. Compared to other approaches, this paper does not goes too much far away. FSM is a well known modeling method and scripting languages are being used for a long time in the context of repetitive activities. The use of both methods together in the same tool enables the conception of a single modeling environment that can be shard during the space product life cycle (from concept to operations).

The idea for the future of these models is to evolve them in order to model the behaviour for different operational plans/situations and being capable of modeling faults at the subsystem level. Some studies are being developed to use the ATOM to simulate the operational procedures of the NanoSatC-BR2, 2U CubeSat developed by INPE to be launched in March/2021. These models will help the Operational Team, at the Ground Stations, to be trained to deal with possible situations that can happen at the in-orbit phases and to analyze the autonomous operation of the payloads on board of the nanosatellite.

\bibliographystyle{IEEEtran}
\bibliography{refe.bib}

\end{document}